\documentclass[a4paper,fleqn]{cas-dc}

\usepackage[numbers]{natbib}

\usepackage{algorithm}
\usepackage{algorithmic}

\usepackage{setspace} 
\usepackage{graphicx} 
\usepackage{amsmath}
\usepackage{array}

\usepackage{url}

\usepackage{breakurl}
\usepackage{hyperref}
\hypersetup{colorlinks,allcolors=black}

\usepackage{array}
\usepackage{graphicx}
\usepackage{booktabs}
\usepackage{mathtools, cuted}
\usepackage{multirow}
\usepackage{xcolor}
\usepackage{subcaption}
\usepackage{todonotes}

\usepackage{setspace}
\usepackage{float} 
\usepackage{url}
\usepackage{xcolor}
\usepackage{soul}
\usepackage{tikz}
\usepackage{multirow}
\usepackage{graphicx}
\usepackage{array}
\usepackage{amsmath}
\usepackage{amsfonts}
\usepackage{amssymb}
\usepackage{listings}
\usepackage{pgfplots} 

\def\tsc#1{\csdef{#1}{\textsc{\lowercase{#1}}\xspace}}
\tsc{WGM}
\tsc{QE}
\tsc{EP}
\tsc{PMS}
\tsc{BEC}
\tsc{DE}

\begin{document}
\let\WriteBookmarks\relax
\def\floatpagepagefraction{1}
\def\textpagefraction{.001}
\shorttitle{Defending SDN against packet injection}
\shortauthors{Phu et~al.}

\title [mode = title]{Defending SDN against packet injection attacks using deep learning}                      



\author[1]{Anh Tuan Phu}
\ead{anhtuan.phu@student.adelaide.edu.au}

\author[1]{Bo Li}
\ead{b.li@adelaide.edu.au}

\author[1]{Faheem Ullah}
\ead{faheem.ullah@adelaide.edu.au}

\author[2]{Tanvir Ul Huque}
\ead{tanvir.huque@yahoo.com}

\author[1]{Ranesh Naha}
\ead{ranesh.naha@adelaide.edu.au}
\cormark[1]

\author[1]{Ali Babar}
\ead{ali.babar@adelaide.edu.au}

\author[1]{Hung Nguyen}
\ead{hung.nguyen@adelaide.edu.au}

\address[1]{The University of Adelaide, Australia}

\address[2]{CyberCX, Australia}

\ead{ranesh.naha@adelaide.edu.au}







\cortext[cor1]{Corresponding author}


\begin{abstract}
The (logically) centralised architecture of the software-defined networks makes them an easy target for packet injection attacks. In these attacks, the attacker injects malicious packets into the SDN network to affect the services and performance of the SDN controller and overflow the capacity of the SDN switches. Such attacks have been shown to ultimately stop the network functioning in real-time, leading to network breakdowns. There have been significant works on detecting and defending against similar DoS attacks in non-SDN networks, but detection and protection techniques for SDN against packet injection attacks are still in their infancy. Furthermore, many of the proposed solutions have been shown to be easily by-passed by simple modifications to the attacking packets or by altering the attacking profile.
In this paper, we develop novel Graph Convolutional Neural Network models and algorithms for grouping network nodes/users into security classes by learning from network data. We start with two simple classes - nodes that engage in suspicious packet injection attacks and nodes that are not. From these classes, we then partition the network into separate segments with different security policies using distributed Ryu controllers in an SDN network. We show in experiments on an emulated SDN that our detection solution outperforms alternative approaches with above 99\% detection accuracy on various types (both old and new) of injection attacks. More importantly, our mitigation solution maintains continuous functions of non-compromised nodes while isolating compromised/suspicious nodes in real-time. All code and data are publicly available for reproducibility of our results.
\end{abstract}








\maketitle

\section{Introduction}
\label{sec:introduction}


Software-defined networking (SDN) improves network performance by enabling easy network management, control and data handling. It is considered a potential solution to problems that are inherent in traditional network architecture and is gaining more popularity in managing virtualised networks, such as cloud infrastructures. SDN, for example, is being employed in data centers and workload optimised systems \cite{sherwood2010carving}. The separation of control plane and data plane gives network administrators ability to manage the network using centralised approach, accelerate provisioning of both physical and virtual network devices. 

The networking architecture of the SDN, however, makes it an easy target for packet injection attacks.  In SDNs, a logically centralised controller collects the devices information (e.g., host location,
switches information) and constructs a global topology view.
Based on the topology view, the controller then deploys
forwarding policies by installing flow rules on each networking
device (e.g., switches). When deployed in an enterprise network, any data packet flow coming from \textit{Outside network}, \emph{i. e.}, other networks, passes through the gateway switch to get access into the network.  Similarly, the data packet flow, generated in the enterprise network and destined to \textit{Outside network}, goes through the gateway switch of the enterprise network. Thus the gateway switch acts as the only access point for the data packet flows to come in$/$go out to$/$from the network which works reactively, whereas the core switches work proactively in the software-defined enterprise network~\cite{2019-CommL-Inspector}.


An OpenFlow switch working in the reactive mode triggers a \textsf{Packet-In} message for any unmatched data packet flow when it encounters and, then  sends the message to the controller.
After getting \textsf{Packet-In} message, the controller installs rules at the  SDN switch to manage that flow~\cite{Sciencedirect-openflow-switch}. 
The controller either allows the data packet flow to pass through the core network or simply drops the data packet flow at the gateway switch. Note that, with the increasing number of the new incoming data packet flows at the gateway switch, the number of \textsf{Packet-In} messages managed by the controller and the number of rules managing the incoming data packet flows installed at the gateway switch increases.

This type of deployment opens the door for a new type of attack - called \emph{Packet Injection Attacks}~\cite{TIFS-2018-packetchecker}. In this attack, the attacker can affect the services and performance of the SDN controller and can overflow the capacity of the SDN switch
devastatingly, by injecting the malicious packets into the SDN network.  Since the SDN controller has no mechanism to verify the legitimacy of \textsf{Packet-In} message, whether it is coming from matched or unmatched packet, adversarial users can exploit this principle of SDN by sending an excessive amount of falsely crafted packets from multiple end hosts to flood the network \cite{deng2017packet}. Doing so ultimately stops the network functioning in realtime, leading to the situation of network breakdown. Thus, the
packet injection attacks is a primary threat to the software-defined enterprise network, in which continuous connectivity and real-time network functioning are two essential requirements.




Detection and defending against these types of attacks are hard.  The recent solution techniques of \textit{Packet~injection} attack, discussed in Section~\ref{sec:related-works}, can stop the malicious data packet flows from entering the network up to a certain level successfully, but these techniques cannot prevent the SDN switch to sending the malicious \textsf{Packet-In} messages to the controller.  Furthermore, more recent low-profile attacks have shown to be able to bypass most of the current detection methods. To address these two open issues in defending SDN against packet injection attacks, we propose in this paper to use a separate detection method that does not involve the controller.  Our solutions continuously monitor traffic in the network, use deep learning models to differentiate between benign and attack traffic. Once an attack is detected and the malicious nodes are identified, the controller installs a set of blocking rules at the switches that allow normal traffic but stop malicious traffic without triggering Packet-In events. The mitigation algorithm scales linearly with the number of switches, the number of hosts and the number of attacking nodes. 

Our main contributions in this paper are
\begin{enumerate}
    \item A novel deep learning model that uses graph convolutional neural networks to detect and identify various types of packet injection attacks in SDN with high accuracy. Our solution is the first solution that can detect and distinguish in real-time between different types of attacks such as slow/fast DoS in SDN.
    \item A scalable SDN-based mitigation solution that helps defend networks against various types of packet injection attacks without disrupting normal traffic.
    \item SDN emulation setup with Mininet that demonstrates the effectiveness of our solutions and generate for the first time a publicly (emulated) dataset of packet injection attacks.
\end{enumerate}

Our code and  data are publicly available at:

\href{https://github.com/nahaUoA/SDN\_PacketInjectionAttack.git}{https://github.com/nahaUoA/SDNPacketInjectionAttack}.

The rest of this paper is organised as follows. We discuss the background information about \emph{Packet Injection Attacks} and their impacts in Section \ref{sec:background-motivation-works}. We then describe the models, hypotheses and techniques used in this work to address limitations in current state-of-the-art packet injection defence in SDNs in Section \ref{sec:proposed-technique}. Section \ref{sec:gcn} explains the two-layer detection using Graph Convolutional Network (GCN) with the proposed attack detection and identification model.  We discuss the performance and results of our techniques in Section \ref{sec:evaluation}. The related works are mentioned and discussed in Section \ref{sec:related-works}. Finally, we conclude our work in Section \ref{sec:conclusion}.


\section{Background and Motivation}
\label{sec:background-motivation-works}

In this section, we describe the motivation of our work by illustrating the effects of the packet injection attack on both the controller and the SDN switches. We also discuss different variants of the packet injection attacks.

\subsection{Packet injection attacks}
\label{subsec:Packet-injection-attack}

By default, an SDN switch in reactive mode  generates and sends the Packet-In message to the controller when it receives an unknown data packet flow~\cite{Sciencedirect-openflow-switch}.
The controller then installs rules at the switch to manage that flow after receiving the Packet-In message. 
The controller either allows the data packet flow to the network or drops the flow at the switch.

The number of Packet-In messages increases in proportion with the number of incoming unknown data packet flows at the switch. The increasing number of Packet-In messages places heavy computational burden on the controller.
Attackers can and do take advantage of  Packet-In messages to launch packet injection attacks on  SDN networks. 
The packet injection attacker sends an enormous number of malicious data packet flows to the switch that slows down or even breakdowns the network.

\subsection{Threat models of packet injection attacks}

Deng\emph{~et~al.}~\cite{TIFS-2018-packetchecker} proposed the initial threat model of the packet injection attack in their seminal work. 
This threat model defines the packet injection attack as a DoS type attack in which the attacker sends numerous malicious packets to an SDN switch ceaselessly.
The majority packet injection attack detection techniques, discussed in Section~\ref{sec:related-works}, use this threat model to validate their proposed solutions.
These techniques identify a network attack as the packet injection attack when the malicious packets coming rate to a switch exceeds a certain threshold for a defined period. 

The threat model of Deng\emph{~et~al.}~\cite{TIFS-2018-packetchecker} cannot detect packet injection attacks when malicious incoming rate to a switch is below the threshold or fluctuates frequently.
In this paper, we address the short comings in Deng's threat model by developing novel threat models that cover a wide variety of packet rates and injection patterns.  More specifically, we  propose two more variants of the packet injection attacks: low-rate packet injection attack and discontinuous packet injection attack. We explain in further detail the two new modes of attack below. Note that it is also possible to have a hybrid type packet injection attack that combines two or more variants of the packet injection attacks.

\subsubsection{Low-rate packet injection attack}

We define the low-rate packet injection attack as a variant of the packet injection attack in which the attacker's packet sending rate is much lower than the threshold value. For example, Khorsandroo~\emph{et~al.}~\cite{Khorsandroo-LCN-2019} shows that an attacker sending malicious packets at a rate of almost $1\%$ of total throughput can exhaust both the switches and the controller significantly. In these cases, the attacker's malicious packet sending rate to a switch does not cross the threshold value, render solutions based on detecting high packet rates ineffective. By launching this attack to multiple switches of a network simultaneously, the attacker can increase the computational burden on the controller to the same level as in  high-rate packet injection attacks.

\subsubsection{Discontinuous packet injection attack}
Another shortcoming of current detection techniques is that they can only detect the malicious high rate attacks over a certain period of time. They fail when  malicious packet coming rate fluctuates around  the threshold value within the fixed detection time window, even if the overall attack volume is very high. Attackers can easily by-pass these detection techniques by following an irregular sending pattern.
We define this type of attack as the \emph{discontinuous packet injection} attack in which the attacker's malicious packet sending rate frequently varies over time.

\subsection{Impact of packet injection attack on  the network}

In packet injection attacks, an attacker intentionally sends a large number of forging data packet flows to the switch to overflow both the switch and the controller capacity. The devastating effect of this attack on the network is further discussed below.

\subsubsection{Computational burden to the controller}

By sending an enormous number of Packet-In messages to the controller, an attacker can keep the controller busy in managing the malicious Packet-In messages. The increasing number of the Packet-In messages increases the controller's CPU utilisation proportionally.
It results in scarcity of resources of the SDN controller managing the regular functionalities of the network.
That eventually stops the controller functioning in real-time, leading to
the enterprise network breakdown.

For example, Deng\emph{~et~al.}~\cite{TIFS-2018-packetchecker} have shown in their experiment that when the controller received the Packet-In message at a rate of $1600$~packets/s, it stopped accepting any new Packet-In message of the network.
In our experiment, we used the same network topology of Deng\emph{~et~al.}~\cite{TIFS-2018-packetchecker} and observed that the controller stops functioning in real-time when the incoming Packet-In message rate at the controller exceeds to $600$~packets/s.

\subsubsection{Rule-space overflow of the SDN switch}

Majority techniques, discussed in Section~\ref{sec:related-works}, mitigate the packet injection attack by installing rules at the switch in which each rule is commonly used to manage a request, such as blocking a specific data packet flow. This type of rule mainly functions by dropping the malicious data packet flows at the switch. 
The number of these rules installed at a switch is proportional to the  number of malicious data packet flows encountered by the switch. Rule space itself, however, can be a target of packet injection attacks.

A typical SDN switch can hold around $55000$ rules at a time~\cite{DSN-2016-Teo}, and it can send or receive around $18000$ requests$/$s at its busiest hour~\cite{2018-NOMS-Madanapalli}.
A switch cannot hold a new rule to manage the new request if its rule space is already occupied. It was shown in~\cite{TIFS-2018-packetchecker} that an attacker can easily launch a packet injection attack by sending malicious data packet flows at a rate of $60000$ requests$/$s to the network.  In this case, the switch's rule space will overflow as its rule space exceeds the capacity, that is, $55000$ rules.
Thus, it is essential to limit the number of installed rules managing the packet injection attacks in the rule-space of the switches.

\section{Proposed Technique}
\label{sec:proposed-technique}

\subsection{Design hypotheses}
\label{subsec:hypotheses}

We assume that, 
\begin{itemize}

\item[1] 
A database server stores the network information as \textit{3 tuple} instance, \emph{i. e.}, $<$\textit{MAC addresses, IP addresses, Switch ID:port no}$>$ in a list called $Network\_List$. The controller gets the full access to the database server using a simple password-based authentication system.

\item[2] The switches work in reactive mode and any data packet flow coming to the switch contains its source and destination address as \textit{3 tuple} instance, \emph{i. e.}, $<$\textit{MAC addresses, IP addresses, Switch ID:port no}$>$ in the network, similar to the \textsf{PacketChecker} technique.

\end{itemize}

\subsection{Defence strategy}
\label{subsec:Defence-strategy}

Based on the flow-based traffic, we identify benign users and attackers with the help of a trained classifier using Graph Convolutional Network (GCN). Section \ref{subsec:attack-detection-technique} discusses a detailed description of the attackers' detection technique. The technique models  the activity flows of the devices using GCN and apply deep learning methods to detect nodes with abnormal behaviours. After detecting the attackers, we will classify the attacks into attacking types such as DDoS and PortScan attacks - two popular packet injection attacks in existing datasets. Note that our solution is capable of identifying other types of attacks, not only these two types. Once we identify and classify the attacks, we mitigate those attacks by adding new blocking rules in the switch. We also maintain another list to observe the activities of the network. Following the blocking rule, the switch will discard incoming packets from the malicious device. The observation list keeps analysing the traffic flow to identify possible attacks.

\subsection{Attack Detection Technique}
\label{subsec:attack-detection-technique}
\begin{figure*}
    \centering
    \includegraphics[width=0.9\textwidth]{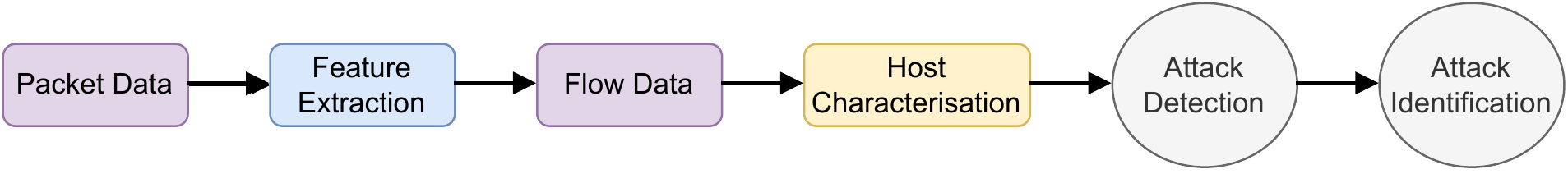}
    \caption{General pipeline of the attack detection and classification architecture}
    \label{fig:pipeline}
\end{figure*}


\subsubsection{Attack Detection}
We used flow-based traffic data to characterise both benign users and attackers' activities. By analysing the traffic generated from each source, we were able to extract the characteristics of the sources and train our classifier to identify the attacking sources. Figure~\ref{fig:pipeline} describes the general pipeline of our packet injection attacks detection and classification.

We first consider all traffic that are not ``benign'' in the dataset to be ``attacks'' traffic. This way, we simplified the problem into binary classification, in which the two classes are ``benign'' and ``attack''. Similar to \cite{almotairi2008characterization}, we group the packets into basic flows; where all flows with the same SrcIP:SrcPort and DstIP:DstPort are grouped together and the mean of their attributes is computed to represent the overall characteristic of the flow. We further group the basic flows into activity flows, where the newly generated basic flows were grouped together based on the source IP and source port. Again, the mean was taken to aggregate the features of those flows. We also attempted to use protocols and count the number of destination ports as features but eventually discarding them because it did not yield a positive result. For nodes that are not associated with any output flows, we perform statistical analysis on the low profile benign sources - gathering the distribution of their features, populates the features of those nodes by sampling from those distributions, and labelling them as ``benign'' nodes. 

By performing statistical analysis on the traffic flows that were transmitted by a source node, we can characterise the time-based features of each node. Attacker nodes tend to have traits that are different from regular source, due to the irregular activities it carries out in the attempts to hijack the network. Finally, we obtained a new representation of data that contains the information on the characteristics of each source, allowing our classifier to learn from those information and perform the predictions.

Using this new representation, we apply  a particular variation of  GCN, called HyperGCN~\cite{yadati2018hypergcn} to identify hosts with malicious packet injection attacks pattern.  HyperGCN is a deep learning model that operates on non-Euclidean structures and allow for multiple connections between two nodes. HyperGCN \cite{yadati2018hypergcn}, which is a variation of GCN that operates on hypergraphs,  is capable of learning structural information and localised information between data samples represented as non-uniform structures, such as graphs, and is suitable for packet injection attacks with multiple flows between switches. Here, we use HyperGCN to solve the problem of network intrusion detection and classification. This problem can be considered as graph-based semi-supervised learning, where the graph represents a network, with users as nodes and connections between users as the edges. Each node is associated with their unique feature vector. Moreover, apart from the node features, GCN also takes the connectivity in the network into consideration. This utilises the power of the graph structure, thus providing useful information for the learning process in case when attackers effectively hide their traits and generate traffic that is indistinguishable from normal traffic. By incorporating the graph structure into the learning process, it is expected that the model can be extremely powerful in scenarios where the links between nodes contain information not present in the data. 




As shown in Figure~\ref{fig:pipeline}, the detection module takes as input the traffic flows. These flows are collected by the mirroring traffic in the network. The detection module produces a list of suspicious packet injection nodes.

\subsubsection{Attack Identification}


Once we detected the malicious hosts, we also want to identify which type of malicious activities the malicious host participate in. 



We extend the GCN algorithms to solve multi-class classification tasks by using them to classify the characterised hosts. A similar approach for representing the data was taken as in the `Attack Detection' stage. In this stage, the algorithms operate only on the malicious samples of the dataset, facilitates the classification process and potentially improve the classification performance.

Using the same graph structure as with the \emph{Attack Detection} model, we perform classification of attack vectors on the set of malicious hosts. In this \emph{Identification} module, semi-supervised learning with GCN is modified to handle the task of multi-class classification. Benign nodes, which are known from the \emph{Detection} module, is deactivated in this stage of the pipeline. Only malicious nodes are used to identify attack variants. Labels based on attack classes, whether DDoS/PortScan for CICIDS data or low-rate/discontinuous for SDN application, are assigned to train portion of the graph model. The GCN algorithms effectively learn from the provided data and propagate the information to identify the attack classes of other hosts in the network.

The output of the \emph{Identification} module is the list of suspicious nodes with their classification labels as DDoS or PortScan. This list is then passed on to the controller for implementing mitigation solution against these malicious nodes.

\subsection{Attack Mitigation Technique}
\label{subsec:attack-mitigation-technique}

The last step in our solution is the mitigation technique that could effective fend off various types of packet injection attacks on SDN without overloading the controller.

\begin{algorithm}
\caption{Attack mitigation algorithm}\label{alg:attack_prevention}  
{
\fontsize{9.5pt}{10pt}\selectfont

\begin{algorithmic}[1]
\renewcommand{\algorithmicrequire}{\textbf{Input:}}
\renewcommand{\algorithmicensure}{\textbf{Output:}}

\REQUIRE \hspace{0.45cm}
$(1)~\mathbb{{S}}$ 
\hspace{1.85cm}
$(2)~\mathbb{E}$ 
\hspace{1.85cm}
$(3)~\mathbb{N}$ \\
\vspace{0.1cm}
\hspace{0.95cm}
$(4)~Packet\_In <Pkt\_In_{src}, Pkt\_In_{dest}>$ \\ 
\hspace{0.85cm}
\vspace{-0.15cm}
\ENSURE
\hspace{0.25cm}
$(1)~Block\_List$ 
\vspace{0.2cm}	 
	 \STATE \textbf{Initialize} $k=1$, where $k \in \mathbb{R}$ 
	 \FORALL {$\mathbb{{S}}$ such that $k \leq K $}  
	 \STATE \textbf{Initialize} $i=1$, where $i \in \mathbb{R}$ 
	 \FORALL {$\mathbb{E}$ such that $i \leq I $}  
		\IF { $ Packet\_In \in \mathsf{E}$ } 
	      \STATE \textbf{Install} \textit{Pkt-blocking} rule at switch $\mathsf{S}_i$
	      \STATE \textbf{Update} $Block\_List \longleftarrow Pkt\_In_{src}$ 
	      \STATE \textbf{Break} 
	    \ELSE
	      \STATE \textbf{Initialize} $j=1$, where $j \in \mathbb{R}$ 
	      \FORALL {$\mathbb{{N}}$ such that $j \leq L $}  
		  \IF {$ Packet\_In \in \mathsf{N}$ and $|Pkt\_In_{dest} \cap \mathsf{N}_j| \neq 3$}
		    \STATE \textbf{Update} $Observing\_List \longleftarrow Packet\_In$ 
	        \STATE \textbf{Break} 
	      \ELSE
	        \STATE \textbf{Continue}  	 
		  \ENDIF
		  \ENDFOR
		\ENDIF
	 \ENDFOR
	 \ENDFOR	 
\end{algorithmic} 
}
\end{algorithm}

\begin{table*}[ht]
	\centering
	\caption{Input parameters of Algorithms~\ref{alg:attack_prevention}.}

	\addtolength{\tabcolsep}{-5pt}
	\begin{center}
	\fontsize{8}{8}\selectfont

	\begin{tabular}{|c |c |}
       \hline
       \textbf{Paramater} & \textbf{Representation} \\[2ex]
       \hline       
       $\mathbb{{S}}$ = The set of all switches & $|\mathbb{{S}_E}|= K$ and $\mathbb{{S}_E}$ = $[\mathsf{S}_1, \mathsf{S}_2, ..\mathsf{S}_K]$ \\[1ex]
       \hline
       $\mathbb{E}$ = The set of all entries of $Observing\_List$ & $|\mathbb{E}|= I$ and $\mathbb{E}$ = $[\mathsf{E}_1, \mathsf{E}_2, ..\mathsf{E}_I]$ \\[1ex]       
       \hline
       $\mathbb{N}$ = The set of all entries of $Network\_List$ & $|\mathbb{N}|= L$ and $\mathbb{N}$ = $[\mathsf{N}_1, \mathsf{N}_2, ..\mathsf{N}_L]$ \\[1ex]       
       \hline
       $Packet\_In_{src} := <src\_IP, src\_MAC,$ & $<$Src IP address, Src MAC address, Src switch-port number$>$ \\       
       $src\_SwitchID:Port\_ID>$ &  \\[1ex]
       \hline
       $Packet\_In_{dest} := <dest\_IP, dest\_MAC,$ & $<$Dest IP address, Dest MAC address , Dest switch-port number$>$\\ 
       $dest\_SwitchID:Port\_ID>$ &  \\[1ex]     
       \hline
	\end{tabular}
	\end{center}
	\label{tab:parameters}
\end{table*}

The controller runs Algorithm~\ref{alg:attack_prevention} recurrently to
mitigate the packet injection attacks at switch level. The inputs of Algorithm~\ref{alg:attack_prevention} is listed in Table~\ref{tab:parameters}.
Algorithm~\ref{alg:attack_prevention} takes inputs from $Observing\_List$ and $Network\_List$ lists, and keeps updating $Block\_List$ list recurrently.
Note that, $Network\_List$ contains the information of all hosts $($benign and suspicious hosts$)$ connected to the network, $Observing\_List$ contains the information of the suspicious hosts, and $Block\_List$ stores the information of the attackers.
These lists keep the host $($benign, suspicious, and attacker$)$ information as $3-$tuple instances, \emph{i.~e.}, $<$IP addresses, MAC addresses, Switch ID$:$port nunmer$>$.

In the SDN network, when a switch encounters an unknown data packet flow, it sends $Packet\_In$ messages to the SDN controller. $Packet\_In$ messages contains the source address, $Packet\_In_{src}$, and the destination address, $Packet\_In_{src}$, of that flow. The controller matches the source address of the received $Packet\_In$ message, $Packet\_In_{dest}$, with the entries of $Observing\_List$.
The controller installs the \textit{Pkt-blocking} rule at the switch to drop the data packet flow immediately (step $2$ to step $8$ in Algorithm~\ref{alg:attack_prevention}), when it gets an exact match between $Packet\_In_{src}$ and an entry of $Observing\_List$.

If the controller gets mismatch between $Packet\_In_{src}$ and an entry of $Observing\_List$, it further matches the destination address of the received $Packet\_In$ message, $Packet\_In_{dest}$, with entries of $Network\_List$.
It adds the sources address of that $Packet\_In$ message, $Packet\_In_{src}$, to the $Observing\_List$ (steps $10$ to $14$ in Algorithm~\ref{alg:attack_prevention}), if it gets a partial mismatch between $Packet\_In_{dest}$ and an entry of $Network\_List$. 
Here the logic is that malicious packets, sent by the packet injection attackers, generally cannot have all fields correctly set (hypothesis) as every single packet must be classified by the switch as ``unknown'' to be forwarded to the controller.
On the other hand, the controller installs a rule at the switch allowing the data packet to flow into the network (step $16$ in Algorithm~\ref{alg:attack_prevention}), when it gets the exact match between $Packet\_In_{dest}$ and an entry of $Network\_List$.

Note that the overall run time of Algorithm~\ref{alg:attack_prevention} is $O(KIL)$ - linear with the numbers of switches, hosts and attackers. 

\section{Two layer detection using GCN}
\label{sec:gcn}

We propose a two-layer detection model, or a dual models architecture to detect whether the coming source of the packets is an attacker and identify which type of attack it is. Furthermore, we also attempt to provide mitigation techniques depending on which type of attack it is, and implement this into the SDN controller.

\subsection{Graph Convolutional Network}

As described in \cite{kipf2016semi}, Graph Convolutional Network (GCN) is a graph-based neural network model, which has the layer-wise propagation rule as presented below
\begin{equation}
    H^{(l+1)} = \sigma(\widetilde{D}^{-\frac{1}{2}} \widetilde{A}\widetilde{D}^{-\frac{1}{2}}H^{(l)} W^{(l)})
\end{equation}

The model f(X, A) is flexible and provides sufficient information propagation across the graph, and is useful for the problem of semi-supervised node classification. On this class of problem, adjacency matrix A and data X of the graph structure is used for training the model. We used a two layer forward model in this network intrusion detection, which is similar to the forward model in \cite{yadati2018hypergcn}:
\begin{equation}
	Z = f(X,A) = softmax(\hat{A}ReLU(\hat{A}XW^{(0)})W^{(1)})
\end{equation}
with
\begin{equation}
	\hat{A} = \hat{D}^{-\frac{1}{2}}\hat{A}\hat{D}^{-\frac{1}{2}}
\end{equation}

The weights are trained using gradient descent. Full dataset is used for every training iteration because the training process actually propagates through the entire graph in each part of training. We also include dropout for stochasticity and weight decay is also used to prevent overfitting.

HyperGCN \cite{yadati2018hypergcn} is a new method that tackles the problem of training GCN with hypergraphs. This method approximates the hypergraph by a graph with each hyperedge is represented by a subgraph with connections of that hyperedge . The connections involve the edge between maximally disparate nodes and edges of these two nodes with all the other nodes in the hyperedge. This operator is called hypergraph Laplacian. After applying hypergraph Laplacian to all the existing hyperedges, an approximation graph is constructed and GCN algorithm is run on this resulting graph.

We apply HyperGCN to solve the problem of network intrusion detection and classification. This problem can be considered as graph-based semi-supervised learning, where the graph represents a network, with users as nodes and connections between users as the edges. Each node is associated with their unique feature vector. By performing statistical analysis on the traffic flows that were transmitted by a source node, we can characterise the time-based features of each node. Attackers nodes tend to have traits that are different from regular source, due to the irregular activities it carries out in the attempts to hijack the network. Moreover, apart from the node features, HGCN also takes the connectivity in the network into consideration. This utilises the power of the graph structure, thus providing useful information for the learning process in case when attackers effectively hide their traits and generate traffics that are indistinguishable from normal traffic. By incorporating the graph structure into the learning process, it is expected that the model can be extremely powerful in scenarios where the links between nodes contain information not present in the data. To the best of our knowledge, our work is the first attempt to apply GCN for network intrusion detection. 
\begin{figure}[htp]
    \centering
    \includegraphics[width=8cm]{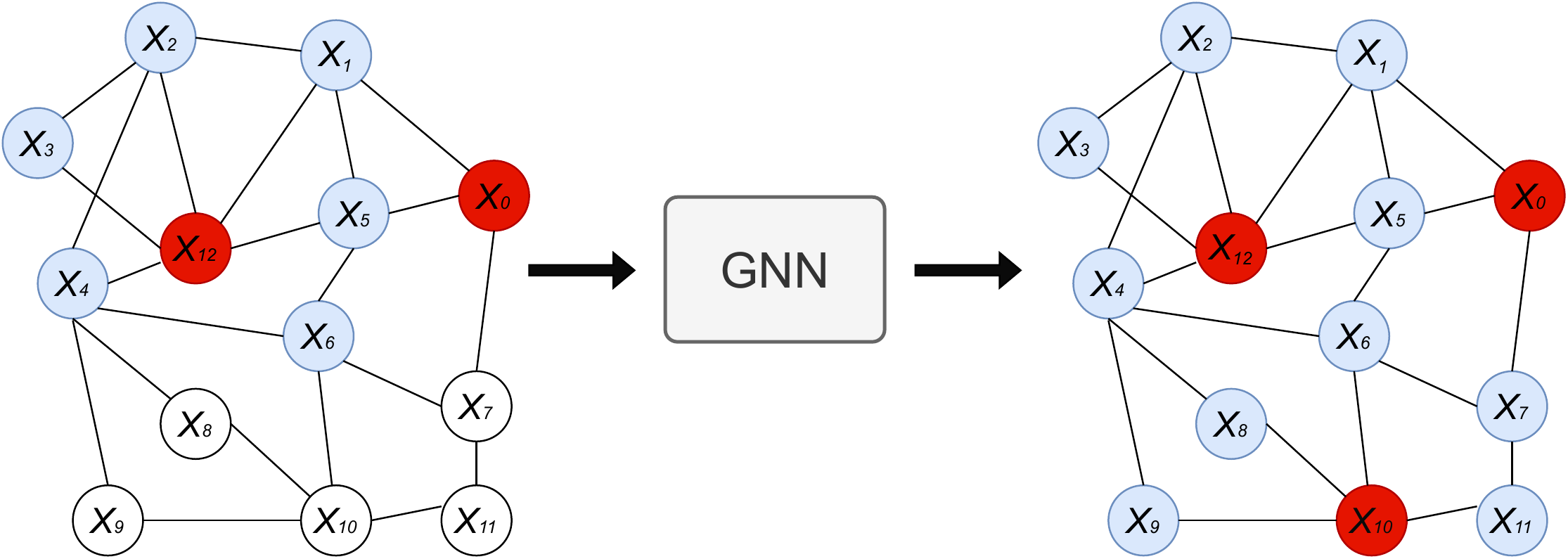}
    \caption{Semi-supervised node classification for intrusion detection with Graph Neural Networks (GNN), blue nodes denote benign users, red nodes denote attackers and white nodes denote unlabelled users. }
    \label{fig:gnn}
\end{figure}

\subsection{Proposed Detection Model}
\begin{figure}[htp]
    \centering
    \includegraphics[width=6cm]{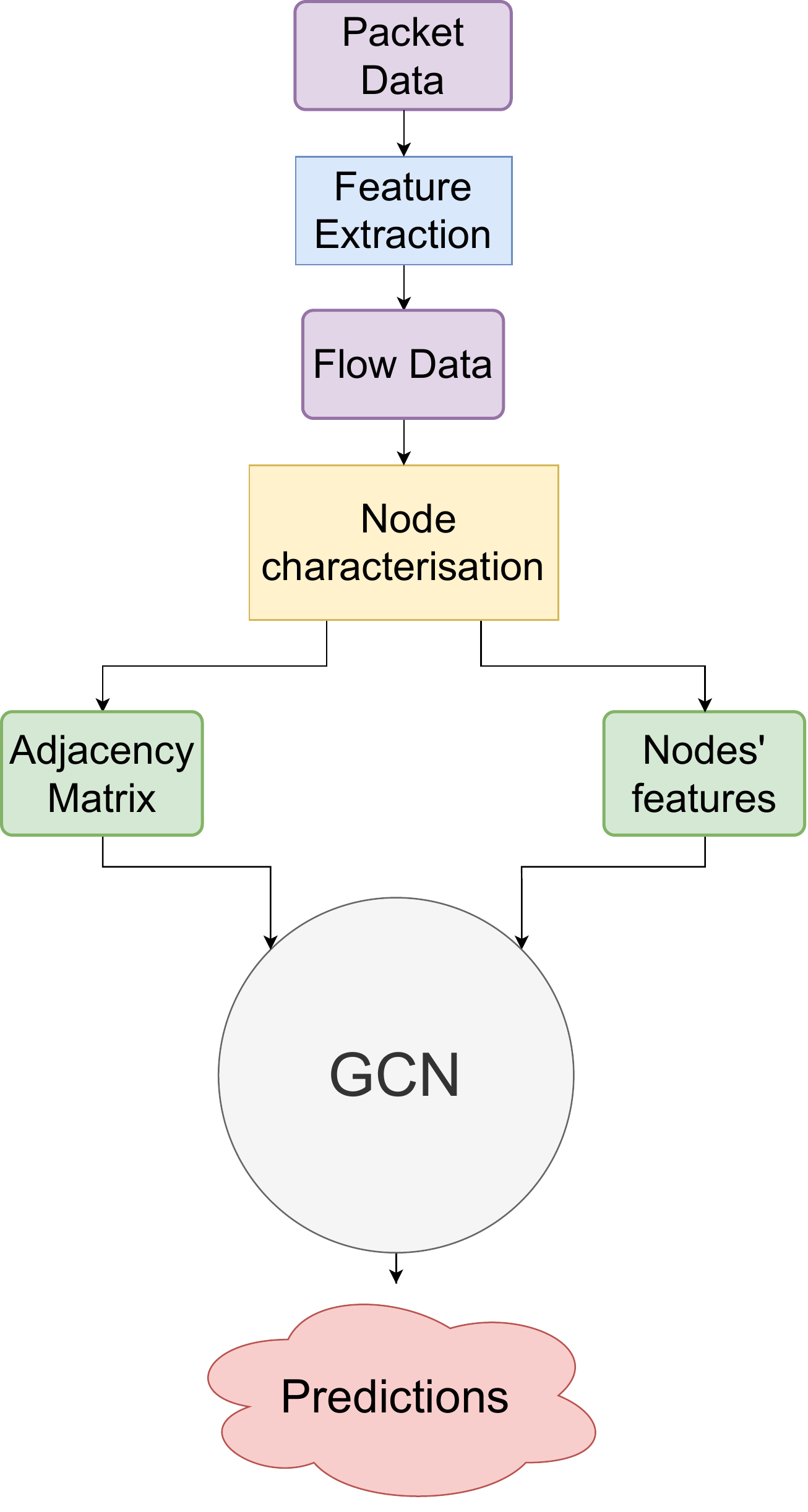}
    \caption{General pipeline of intrusion detection model} 
    \label{fig:pipeline}
\end{figure}

\subsubsection{Layer 1: Attack Detection}
 We used flow-based traffic data (CIC-IDS2017) to characterize both benign users and attackers' activities. Details of this dataset can be found in section VI. By looking at the traffic generated from each source and using some data processing techniques, we were able to extract the characteristics of the sources and train our classifier to identify the attacking sources. 
 
\begin{figure*}[htp]
    \centering
    \includegraphics[width=17cm]{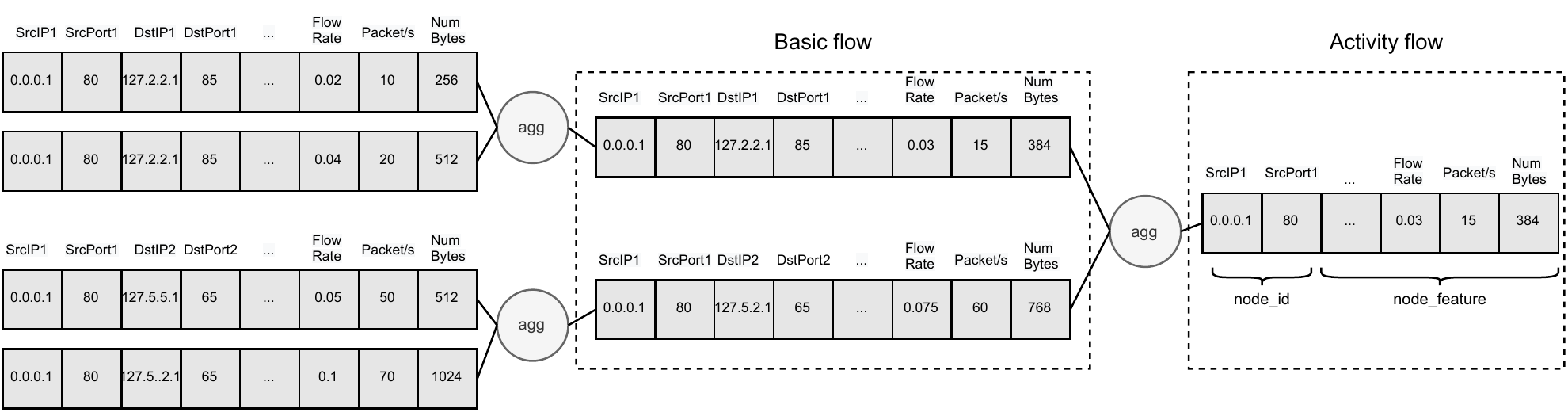}
    \caption{Node Characterization process for extracting nodal features.}
    \label{fig:node_characterisation}
\end{figure*}

 We first consider all traffic that are not ``benign'' in the dataset to be ``attacks" traffic. This way, we simplified the problem into binary classification, in which the two classes are ``benign" and ``attacker". For the GCN method, graph structure needs to be extracted from traffic data. We did this by representing each IP:PORT that sends out or receives flow as a node in the graph, and the edge between nodes exists if there are flows between them. Based on the technique presented in \cite{almotairi2008characterization} techniques, \textcolor{black}{the feature vector of the nodes are obtained using the following steps, or as illustrated in Fig \ref{fig:node_characterisation}:
 \begin{description}
     \item [Step 1:] Group the packets into \emph{basic flows} - all flows with the same SrcIP:SrcPort and DstIP:DstPort are considered in the same group.
     \item [Step 2:] Take the average of attributes, to obtain an overall representation of each \emph{basic flow} object.
     \item [Step 3:] Further group the \emph{basic flows} into \emph{activity flows}, by considering all the objects with the same SrcIP:SrcPort in the same group.
     \item [Step 4:] Use mean aggregation to obtain a feature vector for each \emph{activity flow} object.
     \item [Step 5:] The aggregated \emph{activity flow} is a node in a graph.
 \end{description}
 }
 
 Regarding nodes that are not associated with any output flows, we perform statistical analysis on the low profile (few edges) benign sources - gathering the distribution of their features, populates the features of those nodes by sampling from those distributions, and labelling them as ``benign" nodes. Finally, we obtained a new dataset that contains the information on the characteristics of each source, allowing our classifier to learn from those information and perform the predictions. 
 
 \subsubsection{Layer 2: Attack Identification}
 We used CICIDS2017 dataset for experiments related to attack identification. This dataset contains the most up-to-date kind of attacks that hackers attempt to use for network intrusion. However, as mentioned earlier, we only consider two classes of attacks, which are Port Scan and DDoS attacks, merely because of the number of samples that present in the dataset. Benign traffic and other kinds of attacks except Portscan and DDoS were removed from the dataset in order to keep the aggregation clean and maintain the traits of each type of attack. Furthermore, we also generated a new dataset, which contains different types (e.g., slowDDoS, fastDDoS and discontinuousDDoS) of DDoS attacks that are common in SDN, and perform classification using the same algorithms on this newly generated dataset.

\section{Evaluation}
\label{sec:evaluation}

In this section, we first present the datasets and emulation details. We then report our evaluation results. 

\subsection{Data sets}
We used CICIDS2017 \cite{sharafaldin2018toward} dataset for the evaluation of our proposed solution. We selected CICIDS2017 dataset because it contains benign and numerous up-to-date attack traffic, which was captured by the CIC (Canadian Institute for Cybersecurity) in a format that resembles the real-world data (PCAP). 


CICIDS2017 dataset contains 10 types of attacks - BruteForce, SFTP, SSH, DoS, Heartbleed, slowloris Slowhttptest, Hulk, Infiltration, DDoS and Portscan. We focused mainly on DDoS and PortScan attackers, discarding other types of attacks, and performed classification between these two types of attacks to evaluate our algorithms. The reason these two types of attacks are chosen is twofold - (i) majority of the samples 52\% in the dataset belong to these two classes and (ii) these two are the most commonly reported attacks on SDN \cite{cui2016sd}\cite{boite2017statesec}. Specifically, there are 2,103,072 samples of benign traffic, and 554,375 samples of attacking traffic. Among the attacking traffic sample, there are  128,025 instances of DDoS traffic and 158,804 instances of Portscan traffic, accounted for 52\% of the total instances in the dataset.

\textbf{\textit{DDoS:}} Distributed denial-of-service (DDoS) attack is a malicious attempt to disrupt the normal traffic of a targeted server, service or network by overwhelming the target or its surrounding infrastructure with a flood of Internet traffic.

\textbf{\textit{PortScan:}} A port scan is an attack that sends packets to a range of server port addresses on a host, aiming to explore the host and discover the vulnerability for potential attacks.

In order to do the classification, we firstly performed traffic analysis to extract the activities pattern of each host such as flow duration and flow packets/sec. We then use ML algorithms to classify between benign sources and malicious sources. We further utilise our algorithms to classify different types of attackers, forming a two-layer model that could distinguish between benign users and attackers, as well as type of attackers in the network with high confidence.

\subsection{Mininet Emulation and Data}

\subsubsection{Emulation Setup}
Ubuntu 20.04 LTS was selected as the operating system for the simulation environment. That includes mininet V2.2.2, Open vSwitch (OVS) V2.13.0 and Python 3.8, which are the necessary software to build the emulation environment. In addition, xterm should be installed to simulate hosts, and install it with the following command.
\lstset{
    numbers = left,
    frame = shadowbox,
    xleftmargin = 20pt,
    xrightmargin = 20pt,
    firstnumber = 1,
    stepnumber = 1,
    breaklines,
    columns = flexible,
}

\begin{lstlisting}
sudo apt-get update -y
sudo apt-get install -y xterm
\end{lstlisting}

\begin{figure}[htp]
    \centering
    \includegraphics[width=7cm]{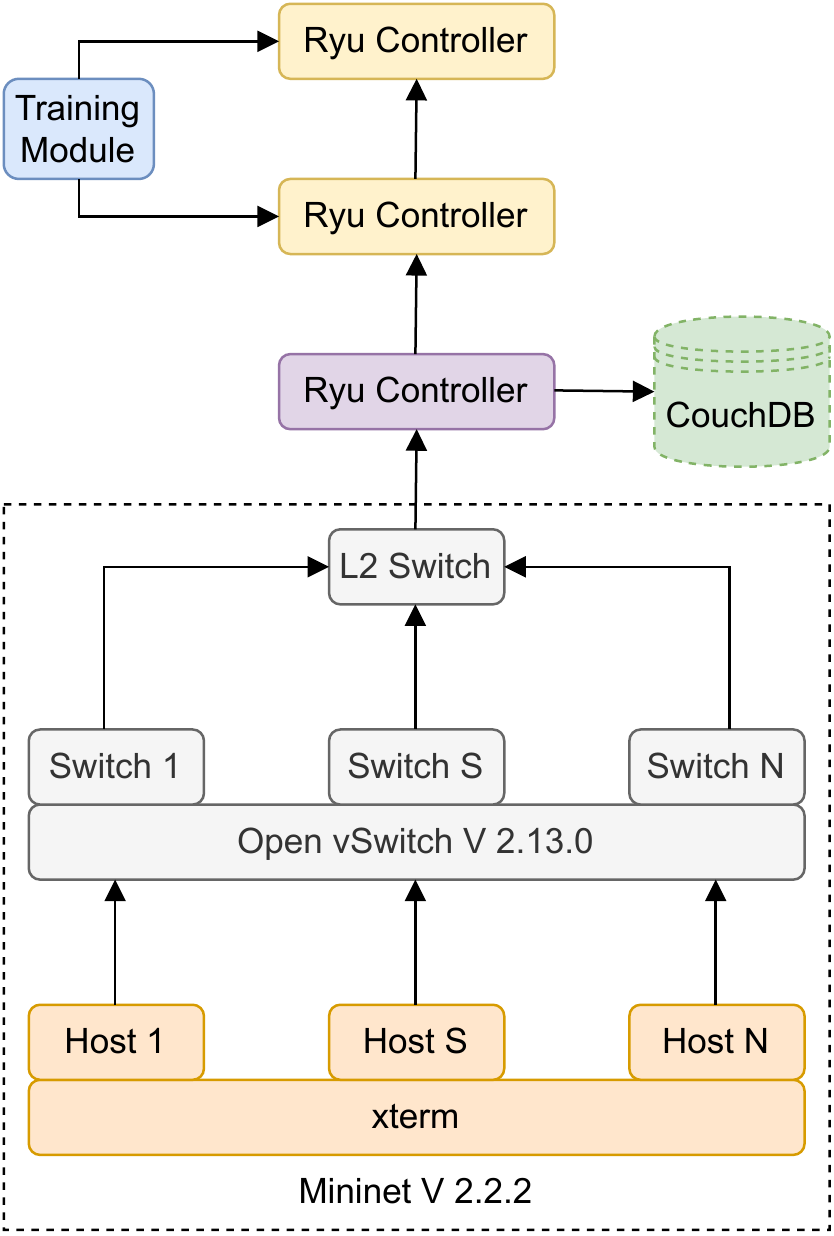}
    \caption{Emulation environment architecture}
    \label{fig:architecture}
\end{figure}

Furthermore, Ryu which is a controller software supporting python also needs to be installed. Use the following command to install, please pay attention to the version of python.
\begin{lstlisting}
sudo apt-get update
sudo apt-get install git -y
git clone git://github.com/osrg/ryu.git
cd ryu/
sudo python ./setup.py install
Finally, CouchDB is installed through subordinate commands.
curl -L https://couchdb.apache.org/repo/bintray-pubkey.asc | sudo apt-key add-
echo ``deb https://apache.bintray.com/couchdb-deb focal main'' | sudo tee -a /etc/apt/sources.list
sudo apt update
sudo apt install couchdb
\end{lstlisting}
After the installation, the mininet emulation architecture is created as shown in Fig. \ref{fig:architecture}. 

\subsubsection{Dataset Generation}
We perform our analyses on two datasets: CICIDS2017 and our own DDoS Variants dataset. The CICIDS2017 is a publicly available dataset that contains samples of benign network traffic and different classes of cyber attacks, including DDoS. We generated the DDoS Variants dataset  using an emulated SDN environment. By varying the various rates and methods of injection, we produced several variants of DDoS attacks in the DDoS Variants dataset, including slow/fast, discontinuous and hybrid attacks.

\begin{figure*}[htp]
    \centering
    \includegraphics[width=13cm]{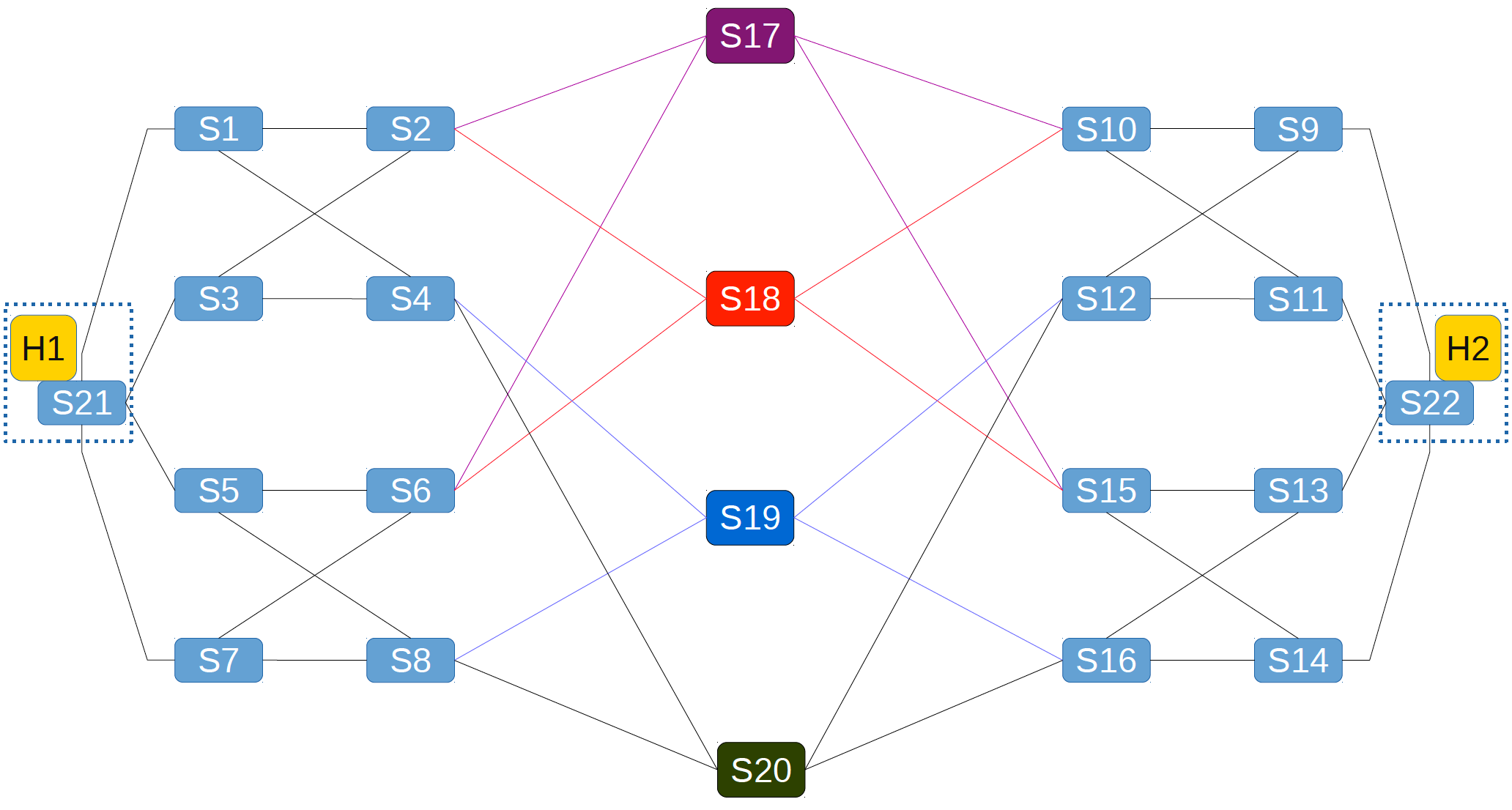}
    \caption{Network Topology}
    \label{fig:nettop}
\end{figure*}

To generate the DDoS Variants dataset, SDN environment were established using Mininet~\cite{OpenFlow-protocol}, with OpenFlow switches and Ryu controller.
Figure \ref{fig:nettop} shows a typical network topology of 2 hosts, using 22 switches that we used in our evaluation. Other topologies could be used without impacting the results of the algorithms.


For packet injection attacks generation, we used \emph{nping} tools to simulate attacking traffic between hosts in the emulated network. Each host was assigned an IP address and carried out different attacks from different port ranges. For the purpose of this work, we simulated four variants of DDoS attacks, namely fastDDoS, slowDDoS, fastDiscontinuousDDoS, slowDiscontinuousDDoS. Table \ref{tab:ddos_variants} shows the property of each DDoS variants. `Rate' parameter is adjusted to vary the rate of DDoS attack and `sleep timeout' parameter is introduced to vary the continuity property. 

In the emulation, we captured data at packet level. We then use CIC-Flowmeter software to extract statistical features of the traffic flow in real-time. A traffic flow can be defined as packets within an arbitrary time period that share the same source/destination IP, source/destination port and transmission protocol. Some of the extracted features include Flow Duration, Number of Packets, Number of bytes, etc.  

\begin{table}
\caption{Properties of DDoS variants}
\begin{tabular}{ | c | c | c | } 
\hline
\textbf{Variant} & \textbf{Rate (packet/s)} & \textbf{Sleep Timeout (s)} \\
\hline
slowDDoS & 5-100 & N/A \\
\hline
fastDDoS & 1000-20000 & N/A \\
\hline
slowDcDDoS & 5-50 & 3-7 \\ 
\hline
fastDcDDoS & 1000-20000 & 3-10 \\ 
\hline
\end{tabular}
\label{tab:ddos_variants}

\end{table}

\subsubsection{Data from Emulation}

After analysing the traffic data, we obtained a new dataset that contains the time-based characteristic data of each source of flow. We stored this in the csv format in order to classify it with other ML techniques. For the HyperGCN algorithm, we used the traffic data to extract the graph. In this graph, each IP:PORT combination that sends out flow to other IP:PORT is considered as a node, and an edge between two nodes is constructed if there is a flow between them. Each node is  associated with features based on their flow pattern, and based on the those features, the HyperGCN, as well as other ML techniques, can identify whether or not they are attackers in the network. And if they are, further classify the type of attacks they involve in.

We used the same approach as in the attack detection problem to characterise the flow pattern of each IP:PORT node. However, some sources involved both Port Scan and DDoS attacks. We addressed this problem by eliminating DDoS traffic and keeping the Portscan. Hence, making the pattern for DDoS and Port Scan be more distinct that is closely aligned to the real-world situation, where a source can only send out one attack type at a time. The reason we chose to eliminate DDoS traffic instead of Port Scan is due to the imbalance of the dataset. By keeping Portscan traffic and discarding DDoS traffic from the sources that send out both, we were able to maintain a more balanced ratio between DDoS and Port Scan traffic in the dataset. For nodes that have no output flow, we assigned them as ``Noflow benign", filling their feature vectors with zeros.

\subsection{Results}
We present results with respect to three parts - attack detection, attack identification, and attack mitigation. 

\subsubsection{Attack Detection}
The HyperGCN model is implemented using PyTorch library. In our implementation, we configured the neural network parameters as in Table \ref{tab:hyperparams} to avoid overfitting. The configuration was obtained via the trial-and-error hyper-parameters tuning method.

\begin{table}
\caption{HyperGCN parameters configuration}
\centering
\begin{tabular}{ | c | c | } 
\hline
\textbf{Params} & \textbf{Values} \\
\hline
\texttt{rate} & \texttt{0.15}  \\
\hline
\texttt{layers} & \texttt{2}  \\
\hline
\texttt{activations}  & \texttt{128}  \\ 
\hline
\texttt{decay} & \texttt{0.0005} \\ 
\hline
\texttt{dropout} & \texttt{0.5} \\
\hline
\end{tabular}
\label{tab:hyperparams}
\end{table}

We compare our HyperGCN-based detection solution against other common machine learning frameworks such as 
\begin{enumerate}

\item
Random Forest (RF)~\cite{muetrj}: RF is a supervised machine learning algorithm that is based on decision trees. It can be considered an ensemble of decision trees that operates on various subsets of the dataset, and the final output is based on the prediction that occurs the most among all decision trees. The RF algorithm was implemented with the number of decision trees set to 100, the maximum depth set to 3 and the random state set to 32. In addition, with the criterion set to 'entropy', 80\% of the datasets were randomly allocated for training and 20\% for testing.

\item
Support Vector Machines (SVM)~\cite{steinwart2008support}: SVM is another classic machine learning algorithm that tailored for feature-rich data. The algorithm constructs a hyperplane or a set of hyperplanes in a high or infinite-dimensional space that separates between classes of samples. For the SVM algorithm implementation, we performed training via SVC, where the kernel parameter was set to 'rbf', C was set to 1, gamma was set to 'auto', and the maximum iteration was set to 8000. In addition, the dataset was also randomly allocated as 80\% for training and 20\% for testing.
\item 
Gradient Boost (GB)~\cite{ke2017lightgbm}: 
Gradient boosting is a form of boosting being used in machine learning. It is based on the assumption that when the best feasible feature model is merged with previous ones, the overall prediction error is minimised. To decrease error, the important aspect to note is to establish the intended outcomes for the next model. The GB algorithm is similar to the RF algorithm in that it is a decision tree-based machine learning algorithm, but the key parameters are different. The n\_estimators and learning\_rate parameters are set to the default values of 100 and 1. The maximum depth parameter is set to 3. The random allocation of datasets is the same as for RF and SVM with 80\% being used for training and 20\% for testing.

\end{enumerate}

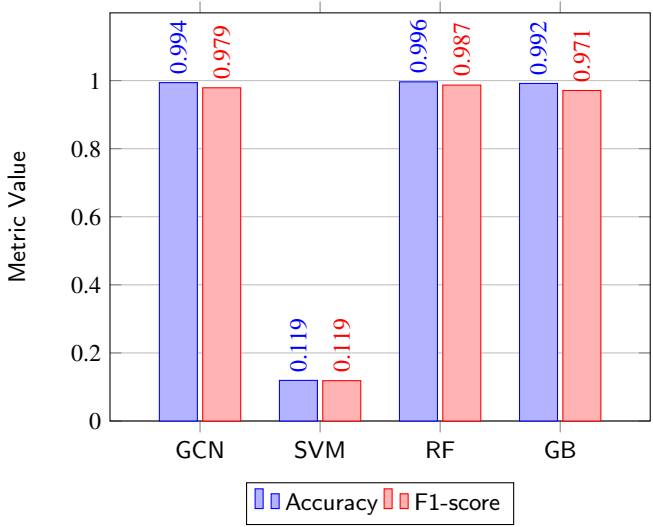
\begin{figure}[h!]
\centering
\pgfplotstableread[row sep=\\,col sep=&]{
    MLalg      & Accuracy & f1score \\
    GCN          & 0.99409  & 0.9789    \\
    SVM          & 0.11933  & 0.11851   \\
    RF & 0.99638  & 0.9868    \\
    GB & 0.99164  & 0.97099 \\
    }\mydatad
    
\begin{tikzpicture}
\centering
    \begin{axis}[
            ybar,
            bar width=.5cm,
            width=0.5\textwidth,
            height=.4\textwidth,
            legend style={at={(0.5,-0.15)},
                anchor=north,legend columns=-1},
            symbolic x coords={GCN,SVM,RF,GB},
            xtick=data,
            ylabel=Metric Value, 
            ymajorgrids = true,
            nodes near coords,
            every node near coord/.append style={rotate=90, anchor=west,/pgf/number format/.cd,precision=3},
            ymin=0,ymax=1.199,
            enlarge x limits=0.25,
        ]
        \addplot table[x=MLalg,y=Accuracy]{\mydatad};
        \addplot table[x=MLalg,y=f1score]{\mydatad};
        \legend{Accuracy, F1-score}
    \end{axis}
\end{tikzpicture}
\caption{Attack detection with CICIDS dataset}
\label{atd}
\end{figure}

CICIDS dataset has the combination of attack and benign. In the dataset, we have to deal with the situation of  imbalanced data, which is common in cybersecurity where the number of benign users accounted for most of the samples (387365 samples), while the number of attackers is much fewer (48396 samples). To avoid overfitting, we selected 20k samples randomly from the benign user class, and 20k samples randomly from the attacker class to construct our training set, and then use the remaining samples for our testing set. Fig. \ref{atd} shows the attack detection results using GCN, SVM, RF and GB machine learning algorithms. SVM underperformed in terms of accuracy and F1-score. However, all three other algorithms (e.g. GCN, RF and GB) attained similar performance for accuracy and F1-score. Furthermore, RF outperformed compared with all other algorithms, and its detection accuracy is close to 100\%.

We also note that even though GCN provides that best overall performance both in term of accuracy and F-1 scores, other ML approaches such as RF and GB have very similar performance. GCN, though, could generalize better than the other approaches as we demonstrate later.

\subsubsection{Attack Identification} We further breakdown the attacker class and classify the type of attacks they are performing, focusing on DDoS and PortScan. This is the second layer of our intrusion detection solution. There is also class imbalance with these classes, with 34254 recorded DDoS attackers and 14142 cases of portscan, the remaining 999 nodes were marked as ``benign noflow''. The GCN model was able to handle this imbalance and with the train/test split of 5 percent training and 95 percent testing on all the samples. We also changed the number of hidden layers to 3 to improve the performance of GCN on this problem. The overall accuracy and F1-score of GCN are 0.9998508 and 0.9995359 respectively.


\textit{Comparisons with other approaches}: Again here, we compare GCN against the performance of other machine learning algorithms, namely  Randomforest, SVM and BoostingTree. In this evaluation study, the same datasets are divided such that 20k attack samples and 20k benign users are randomly selected, and the rest of the data are used for prediction. We referred this data division process as fixed sampling. In addition, based on the same ratio, the datasets division is that 80 percent of the data for training, and 20 percent of the data for prediction. F1-score and accuracy are very important attributes for researching algorithm performance. For imbalanced class data distribution, F1-score is better to evaluate the model because it calculates the harmonic mean of precision and recall.

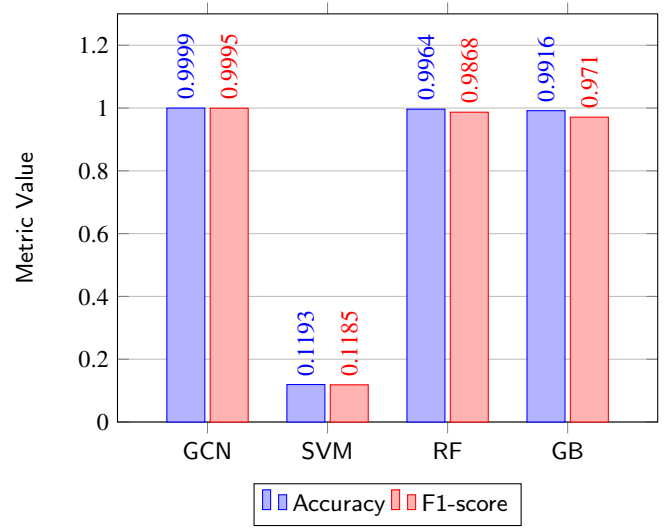
\begin{figure}[h!]
\centering
\pgfplotstableread[row sep=\\,col sep=&]{
    MLalg   & Accuracy & F1score \\
    GCN     & 0.9998508  & 0.9995359    \\
    SVM     & 0.11933455758840275  & 0.1185191109772092   \\
    RF      & 0.9963878581514422  & 0.9868056045444584   \\
    GB      & 0.9916440802552243  & 0.9709918926366381     \\
    }\mydataii

\begin{tikzpicture}
    \begin{axis}[
            ybar,
            bar width=.5cm,
            width=0.5\textwidth,
            height=.4\textwidth,
            legend style={at={(0.5,-0.15)},
                anchor=north,legend columns=-1},
            symbolic x coords={GCN,SVM,RF,GB},
            xtick=data,
            ylabel=Metric Value, 
            ymajorgrids = true,
            nodes near coords,
            every node near coord/.append style={rotate=90, anchor=west,/pgf/number format/.cd,precision=4},
            ymin=0,ymax=1.3,
            enlarge x limits=0.25,
        ]
        \addplot table[x=MLalg,y=Accuracy]{\mydataii};
        \addplot table[x=MLalg,y=F1score]{\mydataii};
        \legend{Accuracy, F1-score}
    \end{axis}
\end{tikzpicture}
\caption{Attack identification with CICIDS dataset with fixed sample}
\label{ati1}
\end{figure} 

Fig. \ref{ati1} shows the results for a fixed sample for different machine learning algorithms. GCN outperformed in terms of accuracy and F1-score SVM, RF and GB. On the other hand, SVM showed worse performance similar to attack detection. Although the attack identification accuracy of GCN, RF and GB are very close to 100\%, if we consider the F1-score, GCN is performing better compared with all other algorithms. Due to the fact of dealing with the imbalanced dataset, a better F1-score proves that the GCN is the best for attack identification.

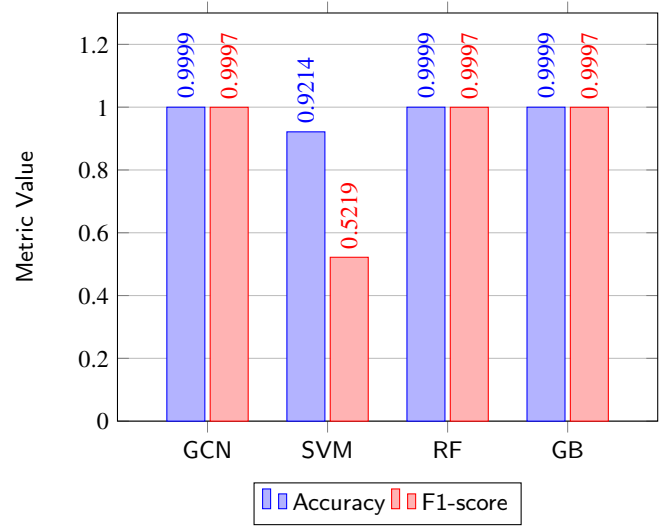
\begin{figure}[h!]
\centering
\pgfplotstableread[row sep=\\,col sep=&]{
    MLalg   & Accuracy & F1score \\
    GCN     & 0.99989  & 0.99966   \\
    SVM     & 0.92144  & 0.52193   \\
    RF      & 0.99989  & 0.99966   \\
    GB      & 0.99989  & 0.99966     \\
    }\mydatai

\begin{tikzpicture}
    \begin{axis}[
            ybar,
            bar width=.5cm,
            width=0.5\textwidth,
            height=.4\textwidth,
            legend style={at={(0.5,-0.15)},
                anchor=north,legend columns=-1},
            symbolic x coords={GCN,SVM,RF,GB},
            xtick=data,
            ylabel=Metric Value, 
            ymajorgrids = true,
            nodes near coords,
            every node near coord/.append style={rotate=90, anchor=west,/pgf/number format/.cd,precision=4},
            ymin=0,ymax=1.3,
            enlarge x limits=0.25,
        ]
        \addplot table[x=MLalg,y=Accuracy]{\mydatai};
        \addplot table[x=MLalg,y=F1score]{\mydatai};
        \legend{Accuracy, F1-score}
    \end{axis}
\end{tikzpicture}
\caption{Attack identification with CICIDS dataset with 80-20 training and testing}
\label{ati2}
\end{figure}















\begin{figure}[h!]
\centering
\pgfplotstableread[row sep=\\,col sep=&]{
    Samples      & GCN         & SVM         & RF          & GB          \\
50   & 0.939356545 & 0.10270992  & 0.958256833 & 0.921434713 \\
250  & 0.983369296 & 0.104578116 & 0.985554529 & 0.971908715 \\
500  & 0.984228081 & 0.106029349 & 0.986891452 & 0.975593055 \\
750  & 0.984441988 & 0.106004123 & 0.987067263 & 0.975174218 \\
1000 & 0.985488075 & 0.105115175 & 0.988010549 & 0.977118933     \\
    }\mydatav

\begin{tikzpicture}
    \begin{axis}[
            ybar,
            bar width=.25cm,
            width=0.5\textwidth,
            height=.4\textwidth,
            legend style={at={(0.5,-0.2)},
                anchor=north,legend columns=-1},
            symbolic x coords={50,250,500,750,1000},
            xtick=data,
            ylabel=Metric Value, 
            ymajorgrids = true,
            nodes near coords,
            every node near coord/.append style={rotate=90, anchor=west,/pgf/number format/.cd,precision=4},
            ymin=0,ymax=1.3,
            enlarge x limits=0.15,
            xlabel={Samples},
        ]
        \addplot table[x=Samples,y=GCN]{\mydatav};
        \addplot table[x=Samples,y=SVM]{\mydatav};
        \addplot table[x=Samples,y=RF]{\mydatav};
        \addplot table[x=Samples,y=GB]{\mydatav};
        \legend{GCN, SVM, RF, GB}
    \end{axis}
\end{tikzpicture}
\caption{Performance when varying number of training samples}
\label{atva}
\end{figure}
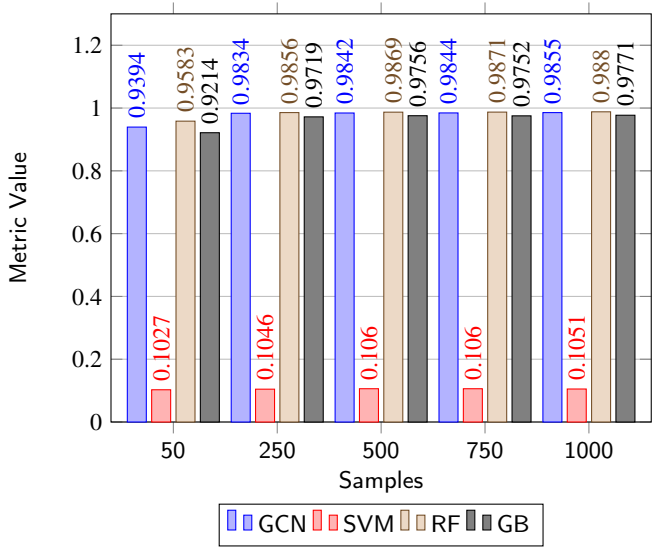  

The experimental outcome of the 80-20 training and testing dataset is shown in Fig. \ref{ati2}. According to the figure, SVM showed  better performance compared with the fixed dataset sample. However, the 0.52 F1-score of SVM makes it unsuitable to use for attack identification. On the other hand, GCN, RF and GB showed the same performance for both accuracy and F1-score. Therefore, we can conclude that fixed sampling is the best for testing the performance and F-score metric is the best for evaluation of attack identification algorithms.




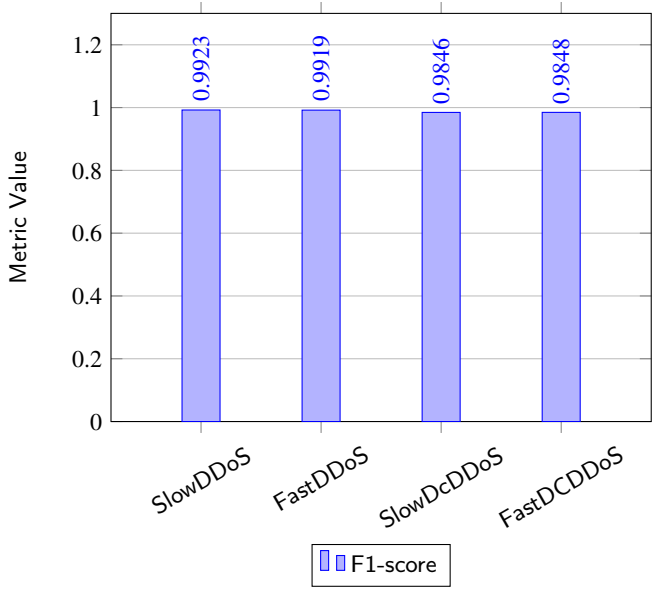
\begin{figure}[h!]
\centering
\pgfplotstableread[row sep=\\,col sep=&]{
    Variant   & F1score \\
    SlowDDoS   & 0.992303585 \\
    FastDDoS     & 0.99186749   \\
    SlowDcDDoS     & 0.984606182   \\
    FastDCDDoS      & 0.984754309   \\
    }\mydatadd

\begin{tikzpicture}
    \begin{axis}[
            ybar,
            bar width=.5cm,
            width=0.5\textwidth,
            height=.4\textwidth,
            legend style={at={(0.5,-0.3)},
                anchor=north,legend columns=-1},
            symbolic x coords={SlowDDoS, FastDDoS, SlowDcDDoS, FastDCDDoS},
            xtick=data,
            ylabel=Metric Value, 
            xticklabel style={rotate=30},
            ymajorgrids = true,
            nodes near coords,
            every node near coord/.append style={rotate=90, anchor=west,/pgf/number format/.cd,precision=4},
            ymin=0,ymax=1.3,
            enlarge x limits=0.25,
        ]
        \addplot table[x=Variant,y=F1score]{\mydatadd};
        \legend{F1-score}
    \end{axis}
\end{tikzpicture}
\caption{Attack identification with DDoS variant dataset}
\label{atddos}
\end{figure}  

We have tested all the algorithms with varying numbers of the training sample for further evaluation. Fig. \ref{atva} shows the test results for F-score with the varying number of training samples. We have varied the test sample size with 50, 250, 500, 750 and 1000 samples for each experiment. The results show that GCN, RF and BB performance are similar and getting better with the increasing number of the training sample. But the performance of SVM is not satisfactory. Testing with 50 samples does not give consistent results compared with  250, 500, 750 and 1000 test samples. From the result, we can say that we need at least 250 test samples to have a consistent F-score. Furthermore, we can conclude that the GCN and RF algorithms have outstanding performance. In addition, although the performance of the RF algorithm is the best among the algorithms used for comparison, the performance of GCN and RF has a very similar performance and the performance difference is only 0.3\%. Hence, we can conclude that we can use either GCN or RF for attack identification. Due to the fact that GCN is computationally expensive compared with RF, we can use RF if we consider computation. However, GCN is great if we do not want to worry much about feature engineering, and it performs better with large datasets.


Following the experimental results from the CICIDS dataset, we have utilised GCN and measured F-score for the DDoS variants dataset. Fig. \ref{atddos} shows F1-score results for SlowDDoS, FastDDoS, SlowDcDDoS and FastDcDDoS variant dataset. For all cases, GCN showed better performance as expected. As shown in Fig. \ref{atddos}, the F-score is between 0.98 to 0.99, which is excellent for an imbalanced dataset. Furthermore, the overall attack identification accuracy is 0.99.

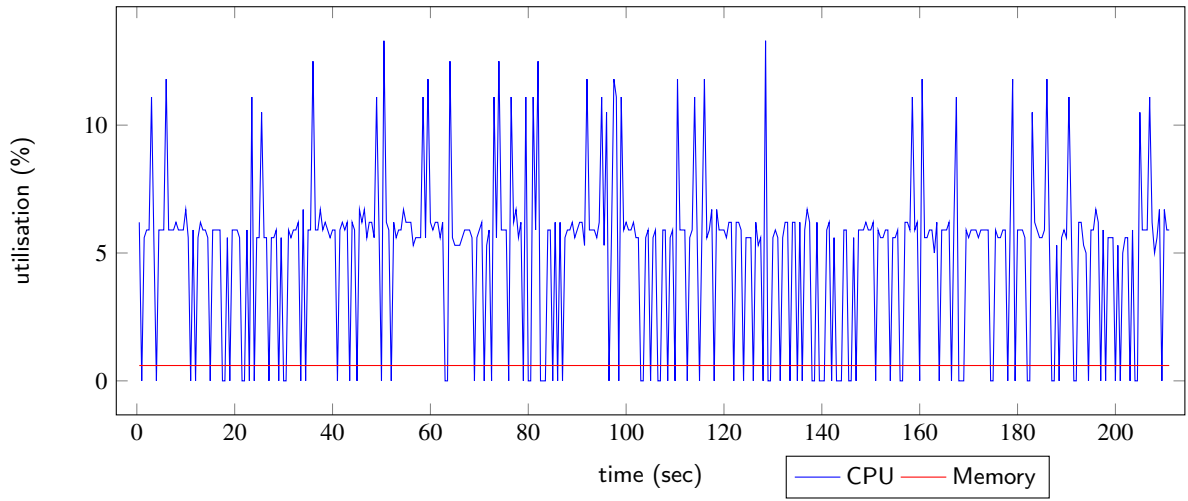
\begin{figure*}[h!]
\centering
\begin{tikzpicture}
\begin{axis} [xlabel=time (sec), 
                ylabel=utilisation (\%), 
                legend style={at={(0.75,-0.1)},
                anchor=north,legend columns=-1},
                width=0.9\textwidth,
                height=0.4\textwidth,
                xmin=0,xmax=210,
                enlarge x limits=0.02]
\addplot [blue, mark=none] table [x=T, y=CPU, col sep=comma] {IdleSleep05s.csv};
\addplot [red, mark=none] table [x=T, y=Memory, col sep=comma] {IdleSleep05s.csv};
\legend{CPU, Memory}
\end{axis}
\end{tikzpicture}
\caption{Idle mode CPU and memory utilisation in every 0.5 second}
\label{ati05}
\end{figure*}

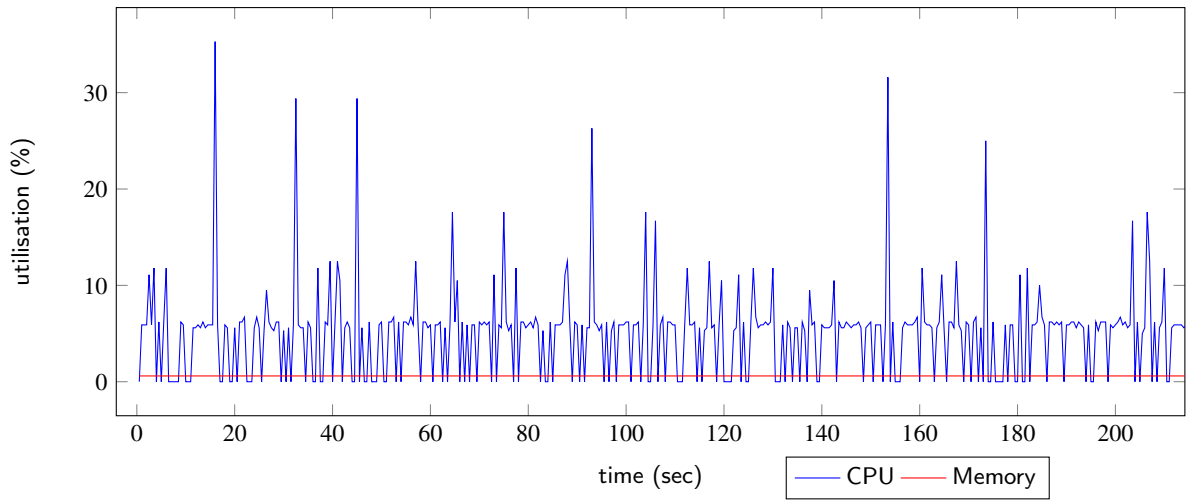
\begin{figure*}[h!]
\centering
\begin{tikzpicture}
\begin{axis} [xlabel=time (sec), 
                ylabel=utilisation (\%), 
                legend style={at={(0.75,-0.1)},
                anchor=north,legend columns=-1},
                width=0.9\textwidth,
                height=0.4\textwidth,
                xmin=0,xmax=210,
                enlarge x limits=0.02]
\addplot [blue, mark=none] table [x=T, y=CPU, col sep=comma] {rate100to500ddosSleep05s.csv};
\addplot [red, mark=none] table [x=T, y=Memory, col sep=comma] {rate100to500ddosSleep05s.csv};
\legend{CPU, Memory}
\end{axis}
\end{tikzpicture}
\caption{CPU and memory utilisation in every 0.5 second with mitigation technique}
\label{atm05}
\end{figure*}

\subsection{Attack Mitigation}

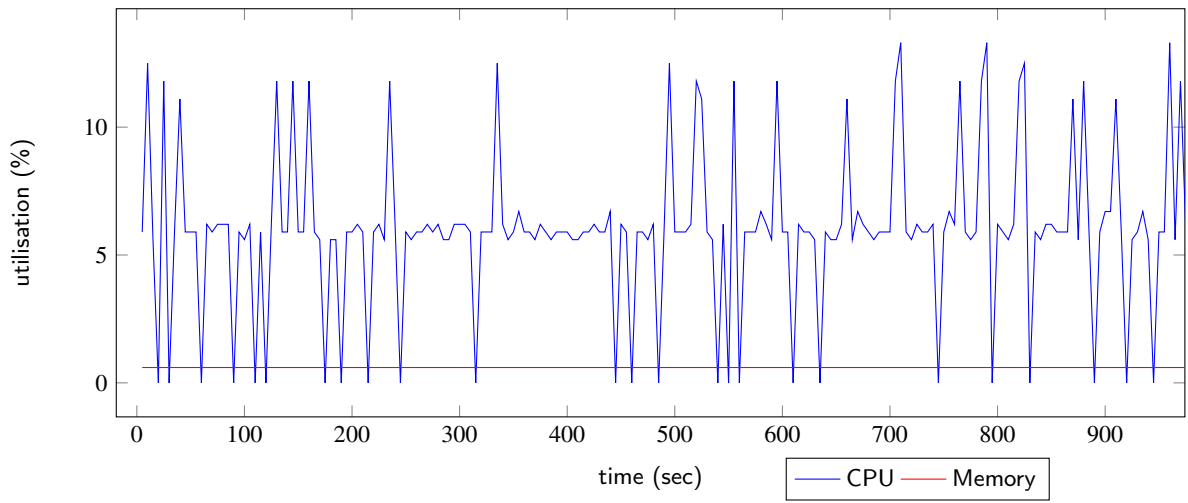
\begin{figure*}[h!]
\centering
\begin{tikzpicture}
\begin{axis} [xlabel=time (sec), 
                ylabel=utilisation (\%), 
                legend style={at={(0.75,-0.1)},
                anchor=north,legend columns=-1},
                width=0.9\textwidth,
                height=0.4\textwidth,
                xmin=0,xmax=955,
                enlarge x limits=0.02]
\addplot [blue, mark=none] table [x=T, y=CPU, col sep=comma] {IdleSleep5s.csv};
\addplot [red, mark=none] table [x=T, y=Memory, col sep=comma] {IdleSleep5s.csv};
\legend{CPU, Memory}
\end{axis}
\end{tikzpicture}
\caption{Idle mode CPU and memory utilisation in every 5 second}
\label{ati5}
\end{figure*}
\begin{figure*}[h!]
\centering
\begin{tikzpicture}
\begin{axis} [xlabel=time (sec), 
                ylabel=utilisation (\%), 
                legend style={at={(0.75,-0.1)},
                anchor=north,legend columns=-1},
                width=0.9\textwidth,
                height=0.4\textwidth,
                xmin=0,xmax=955,
                enlarge x limits=0.02]
\addplot [blue, mark=none] table [x=T, y=CPU, col sep=comma] {rate100to500ddosSleep5s.csv};
\addplot [red, mark=none] table [x=T, y=Memory, col sep=comma] {rate100to500ddosSleep5s.csv};
\legend{CPU, Memory}
\end{axis}
\end{tikzpicture}
\caption{CPU and memory utilisation in every 5 second with mitigation technique}
\label{atm5}
\end{figure*}
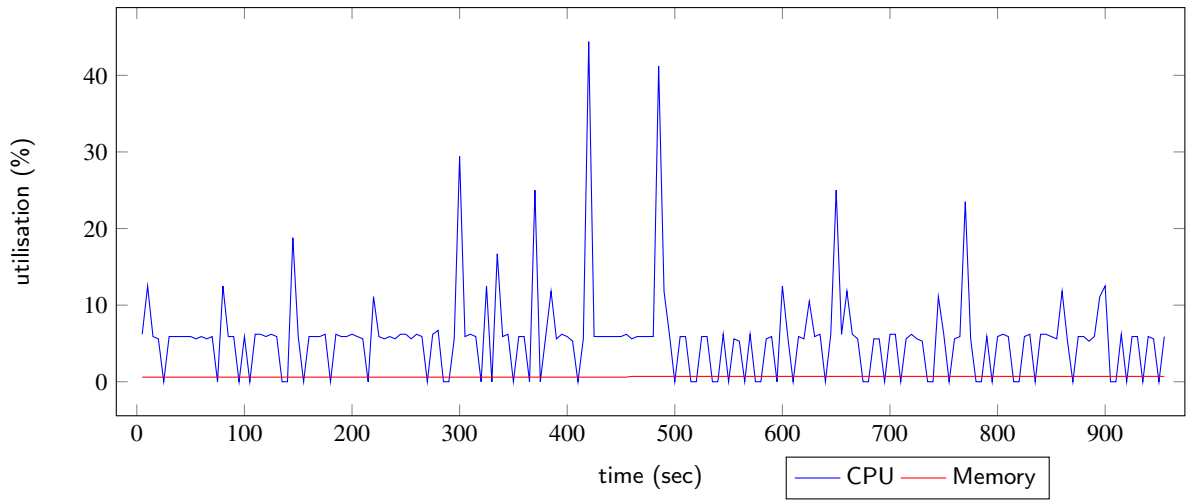

In the proposed defence mechanism, identifying an attack is very important because we can quickly implement the rules in the switch through the controller once the attack is identified. However, the process should not consume too much resources by which the controller might collapse. Therefore, we need to prove that CPU and memory utilisation is less while the attack mitigation technique is running on the controller. We have tested the mitigation technique in four different scenarios: (i) observe CPU and memory utilisation every 0.5 second while the system is in idle mode (ii) observe CPU and memory utilisation every 0.5 second while the system is employing the mitigation technique (iii) observe CPU and memory utilisation every 5 seconds while the system is in idle mode (iv) observe CPU and memory utilisation every 5 seconds while the system is employing the mitigation technique. We have initiated the DDoS attack with a packet sending rate of 100 to 500 in a second. We have logged the utilisation in every 0.5 and 5 seconds to understand the resource utilisation behaviour of the controller. The experimental results from these four scenarios are shown in Figures \ref{ati05}, \ref{atm05}, \ref{ati5} and \ref{atm5}.

Fig. \ref{ati05} shows the utilisation of CPU and memory in idle mode with 0.5 sec utilisation capturing interval. According to the figure, the maximum CPU utilisation is 13.3\%, and the minimum value for CPU utilisation is 0. The mean value and standard deviation are 4.93 and 3.19, suggesting a lot of variation in CPU utilisation over time. However, CPU utilisation does not exceed 13.3\% in idle mode. After applying the mitigation technique in the same scenario with a DDoS attack, CPU utilisation increased to a maximum of 35.3\% (see Figure \ref{atm05}). But the mean value and standard deviation are 5 and 4.72, which suggests CPU utilisation variation is high. However, reaching to 35.3\% CPU utilisation is not that frequent. We did not find any difference in memory utilisation, memory utilisation remains the same over time. It is clear that CPU and memory utilisation do not significantly differ from our proposed mitigation technique. During the mitigation, if resource utilisation reaches to peak, then the method cannot be used in the production environment. But our experimental results prove that the effectiveness of our proposed mitigation technique performs better with a lower resource utilisation.

We experimented with a 5 sec utilisation capturing interval for a longer time than the first scenario with the same packet sending rate. Fig. \ref{ati5} shows the utilisation of CPU and memory in idle mode with 5 sec utilisation capturing interval. According to the figure, the maximum CPU utilisation is 13.3\%, and the minimum value for CPU utilisation is 0, which was exactly the same as the first scenario. The mean value and standard deviation are 5.96 and 3.02, suggesting a similar CPU utilisation with scenario (i). After applying the mitigation technique with the 5 sec utilisation capturing interval with a DDoS attack, the maximum CPU utilisation reached to 44.4\%. But the mean value and standard deviation are 5.72 and 5.97, which suggests a slight increase in CPU utilisation. On the other hand, memory utilisation also increased a little, reached to 0.7 from 0.6 for around 50\% cases. Overall, from our experimental results, we can conclude that the proposed mitigation technique performs better with reasonable resource utilisation.


\section{Related works}
\label{sec:related-works}

We cover in this section related research in both SDN packet injection attacks,  mitigation techniques and machine learning applications in network security.  

\subsection{Packet Injection Attacks on SDN}
\label{subsubsec:pkt-attack-to-SDN}

Packet injection attacks on SDN networks were first discussed in the seminal work of Deng\emph{~et~al.}~\cite{TIFS-2018-packetchecker}.
Their proposed technique, PacketChecker, can detect and mitigate packet injection attacks at switches.
However, this technique increases the computational burden on the network (controller and switches) that the controller may breakdown eventually.
It also increases the rule-space overheads of the switch during the attack.

Alshra'a\emph{~et~al.}~\cite{2019-CommL-Inspector} propose a packet injection attack mitigation technique to keep the network functioning in real-time.
In their approach, a hardware-software device, INSPECTOR, keeps the controller separated from managing Packet-In messages of the malicious data packet flows. 
This technique reduces the computational burden on the controller and rule-space overheads of the SDN switches noticeably.
The primary drawback of their approach is that this technique does not match with the conformity of the SDN architecture.

Both PacketChecker and INSPECTOR techniques need the controller's involvement to tackle packet injection attacks. 
Scott-Hayward\emph{~et~al.}~\cite{Scott-Hayward-sdn-nfv-2018} propose a technique to mitigate packet injection attacks without the controller's intervention. 
This technique uses a stateful data plane approach to stop packet injection attacks at the switch-level in a limited scope. 
This switch-level technique reduces the controller's computational burden and rule-space overhead of switches compared to PacketChecker and INSPECTOR. However, this technique cannot detect the attack when the attacker launches the attack using an unregistered host address.

The above techniques focus only on high-rate packet injection attacks, \emph{i.~e.}, malicious packets coming in at such a high rate that they take out the controller. However, low-rate packet injection attacks, \emph{e.~g.}, $10$ malicious packets$/$s, may also pose a real threat. 
Khorsandroo~\emph{et~al.}~\cite{Khorsandroo-LCN-2019} discuss the effect of low-rate packet injection attacks on the SDN networks but do not offer any solution to mitigate this type of attack.
They show that an attacker generating offensive traffic at a rate of almost $1\%$ of total throughput in off-peak hours can exhaust almost $25\%$ of rule-space overheads of switches and decrease throughput by nearly $40\%$.

Zhan~\emph{et~al.}~\cite{Xinyu-Springer-lecture-note-2020} propose a probabilistic technique to detect and mitigate packet injection attacks to the SDN network. 
They use the machine learning technique to detect packet injection attacks that offer almost $91\%$ attack detection accuracy. Our interest in this paper focuses on designing a deterministic technique to detect and mitigate packet injection attacks. More recently, Tanvir and Frank \cite{ul2021protecting} presented a workload offloading approach for protecting SDN from packet injection attacks. According to this approach, when the malicious traffic inflow crosses a certain threshold, the edge controller takes control from the core controller. This way the core controller remains available even when the gateway switch is under attack. Unlike \cite{ul2021protecting}, our work uses deep learning to detect and mitigate the attack. 

To detect packet injection attacks on SDN, Jishuai et al. \cite{li2022packet} added a new component (PIEDefender) to SDN controller. This add-on is protocol independent and also does not require any additional hardware. Basically, this add-on monitors the verification of \textit{Packet-In} messages. The authors evaluated this add-on using Floodlight controller. It was found that this add-on enables the controller to detect packet injection attacks with 97.8\% precision. Unlike \cite{li2022packet}, our approach does not add any component to the controller but incorporates deep learning to detect packet injection attacks.


\subsection{Deep Learning Methods for Protecting SDN}
\label{subsubsec:deep-learning-methods}

Several studies (e.g., \cite{niyaz2016deep}\cite{tang2016deep}\cite{li2018detection}\cite{garg2019hybrid}\cite{maeda2019botnet}\cite{mehdi2011revisiting}) have used Deep Learning (DL) methods for detecting and preventing various types of attacks on SDN. Niyaz et al. \cite{niyaz2016deep} proposed a DL-based technique for detecting packet injection attacks in an SDN environment. The stacked autoencoder-based DL technique was deployed on SDN controller for extracting useful features from the network traffic. The approach was evaluated using a testbed consisting of an SDN controller (POX), Open vSwitch, 12 network devices, and various attacks launched using hpring3. The results showed that the DL-based approach can classify benign and malicious traffic with 99.82\% accuracy and different types of attacks with 95.65\% accuracy.

Tang et a. \cite{tang2016deep} also proposed a Deep Neural Network (DNN) model for detecting attacks in an SDN environment. The proposed technique relies on a very small number of features (only six) to train the DNN model. The DNN model is deployed on the SDN controller, which receives information of the entire network from the OpenFlow switches for analysing to detect attacks. The authors evaluated the model using NSL-KDD dataset. The results reveal that the model can detect attacks in an SDN environment with 75.75\% accuracy.

The authors of \cite{li2018detection} integrated several DL algorithms (i.e., RNN, CNN, and long short-term memory) to build an attack detection system for detecting attacks on OpenFlow-based SDN. For training the DL models, ISCX dataset was used while the test was conducted via real-time packet injection attacks. Attacks were detected with an accuracy level as high as 98\%. High detection accuracy, minimal dependence on hardware, and easily updating the model were identified as the key advantages of the DL-based attack detection technique.

In order to enhance the reliability of SDN, Garg et al. \cite{garg2019hybrid} proposed a DL-based anomaly detection technique to detect suspicious data flows for social media domain. This hybrid technique integrates Boltzmann machine with SVM for anomaly detection and leverages an end-to-end data delivery approach to satisfy QoS requirements. The proposed approach was evaluated both with benchmark dataset (KDD) and real-time data collected from TIET institute in India. Experimental results reveal upto 99\% anomaly detection rate.

Motivated by the fact that infected nodes can be isolated via SDN, Maeda et al. \cite{maeda2019botnet} proposed MLP's DL-based botnet detection technique that runs on the SDN controller. This way once a botnet is detected, the infected areas (e.g., nodes) of the network are isolated to ensure that the infection does not spread to the rest of the network. For evaluating the approach, the authors used Ryu SDN framework as SDN controller and Open vSwitch as the OpenFlow switch. CTU-12 and ISOT datasets were used to assess the accuracy of the DL technique. It was found that the approach can accurately detection 99.2\% of botnet attacks.

Mehdi et al. \cite{mehdi2011revisiting} demonstrated the implementation of four DL algorithms in NOX SDN controller for anomaly detection. The DL algorithms were evaluated with data collected in three scenarios - ISP router, switch of a lab, and home network router. It was revealed that the anomaly detection technique can accurately detect anomalies in home and lab scenarios, however, fails in the ISP router case. It was also found that the detection technique not only has decent accuracy but also does not significantly slows down the flow of packets in any way.

Majority of the DL-based methods for protecting SDN relies on solely network features, without taking the relational architecture of the network itself into account. In reality, there are many applications where data can be represented in the form of graphs \cite{wu2020comprehensive}. Computer networks can be naturally represented with a graph structure, where nodes can be network objects (i.e., hosts, switches, servers) and the edges can be the connection between those objects (i.e, whether a host talks to a server). Our work successfully captures these relational features between network objects by constructing a graph from network flow vectors and train the GCN model using that graph. Furthermore, this work also leverages the network-flow features for embedding into the nodal feature vector, which will be learnt by the GCN model in addition to the relational features. Aside from the GCN model, our work also differs from the previous works in terms of goals (detection, identification, and mitigation), architecture, datasets, and the evaluation outcomes.


\section{Conclusion}
\label{sec:conclusion}

Detecting and preventing packet injection attacks on SDN in real-time is a challenging task. Therefore, in this paper, we have developed Deep Learning (DL) and Machine Learning (ML) based approaches to secure SDN against packet injection attacks. Our approach first uses DL/ML to detect attacks in real-time. It then identifies the type of attack such as slow/fast packet injection attack. Finally, our approach mitigates the attack by updating the network rules in real-time. We have evaluated our approach using an SDN emulation setup. In the evaluation, we have used two datasets - CICIDS2017 and our own generated dataset, which is the first dataset for packet injection attacks. Our evaluation reveals that (i) Graph Convolutional Network (GCN) and Random Forest (RF) attained the highest accuracy for attack detection and identification (ii) the accuracy of GCN, RF, and Gradient Boosting (GB) improves with the increase in the number of training samples (iii)  all tested algorithms can accurately detect the various types of packet injection attacks (mean accuracy of around 0.99) (iv) our approach consumes minimal resources (CPU and memory) for mitigating packet injection attacks. Our code and data are publicly available at: \href{https://github.com/nahaUoA/SDN\_PacketInjectionAttack.git}{https://github.com/nahaUoA/SDNPacketInjectionAttack}


In future, we plan to investigate the impact of the increasing SDN packet injection attacks in a distributed manner (attack generation from 10 to 100 hosts at least) on the effectiveness of our approach. We also plan to observe the impact on the resource utilization of SDN switches along with the controller as we increase the number of SDN packet injection attacks. Furthermore, we also plan to evaluate our approach in a real testbed environment. Another fruitful direction is to investigate the impact of adding/removing more and more rules to the switches and controllers.

\section*{Acknowledgment}
Tanvir ul Huque acknowledges the support of the Commonwealth of Australia and Cybersecurity Research Centre Limited, Australia.




\bibliographystyle{cas-model2-names}

\bibliography{cas-refs}





\end{document}